# Basic requirements for proven-in-use arguments

Hendrik Schäbe, TÜV Rheinland

Jens Braband, Siemens AG

## Introduction

Proven-in-use arguments are needed when pre-developed products with an in-service history are to be used in different environments than those they were originally developed for. Particular cases may include

- commercial-off-the-shelf (COTS) components that are to be used in safety contexts
- pre-developed products that have been certified by a different standard or against different requirements
- products that have originally been certified to a lower level of integrity but are to be used for higher levels of integrity

A product may include software modules or may be stand-alone integrated hardware and software modules.

IEC 61508-7 [1] includes as Annex D a proposal for the evaluation of software reliability based on a simple statistical model. However, this annex is discussed quite controversially and is currently under revision. EN 50129 [2] also contains recommendations on how to demonstrate high confidence by use, but without any further explanation. For example, for SIL 3 and 4, "1 million hours operation time, at least 2 years experience with different equipments including safety analysis, detailed documentation also of minor changes during operation time" is recommended.

The topic itself is not new [3, 4], but most recent approaches have been based on elementary probability such as urn models which lead to very restrictive requirements for the system or software to which it has been applied [7]. We have only become aware during the process of writing this paper that much earlier work has been performed with a different approach [15].

Also, there has been considerable confusion about the term "software reliability". Therefore, the authors want to state very clearly: software does not fail, only physical systems can fail. So, unless software is executed on hardware, we cannot refer to failure.

The aim of this paper is to base the argumentation on a general mathematical model, so that the results can be applied to a very general class of products without unnecessary limitations. The advantage of such an approach is also that the same requirements would hold for a broad class of products.

## Mathematical model

In this section, we provide a simple model of a pre-developed product. It consists of software and hardware. Hardware is relatively easy to consider. It exhibits random failures and systematic failures. As for random failures, it is necessary to deal with aging, wear and tear, and comparable effects that lead to a possibly dangerous failure of the hardware. Since there are a large number of influences such as material, load, use etc., the failure time is random. This is modeled by distributions such as exponential distribution, Weibull distribution and many other models, e.g. see [10].

Furthermore, the hardware may contain systematic faults. We will treat those together with software errors. Normally, if a component has been developed according to a standard for functional safety, it is assumed that such methods have been used that systematic failures can be neglected, compared with random failures. However, for pre-developed components, this assumption might not hold. Therefore, we need to consider systematic failures of the hardware.

For software, we will now use the following model. In order to simplify the situation, we will assume that there are no randomization procedures in the software, as reading system time, using uninitialized variables, and using data in memory cells and other data that might result in a random influence on the software. These processes, also chaotic processes used for example for random number generation, could be discussed separately.

With these assumptions, the software could be imagined as a deterministic system. This system adopts a new state for each processor step. Obviously, the state of the software would depend on the previous states of the software. Here, "state" means the values of all registers, counters, pointers etc. This means we do not assume that the software fulfils the Markov property. In that case, the next state of the software would only depend on the previous step and not on the history. Now, we need to take into account the fact that the state of the software would also depend on the input data.

Hence, we have a space of input values E and a space of output values A. We can express the input values as a sequence $E_i$, i=1,... Now, the software provides a deterministic mapping $A_n = F(E_n, E_{n-1},..., E_i,...,E_1)$. This mapping is subjective. Now, there exist n+1 tuples ($A_n$, $E_n$, $E_{n-1}$,..., $E_i$,...,$E_1$) that are either correct (C) or incorrect (N). This is just a reflection of the software errors that are present. So, the entire space ($A_n$, $E_n$, $E_{n-1}$,..., $E_i$,...,$E_1$) for all n is subdivided into two parts: C and N. Usually, C would have a much larger size than N. Size here means the number of points. It needs to be understood that there would be a very large number of points in a space. That means, if we chose at random a point from the set of all possible ($A_n$, $E_n$, $E_{n-1}$,..., $E_i$,...,$E_1$), we would only rarely be in N.

Inputs E1, E2, etc., can be considered as random. However, their distribution influences how often the software status would belong to subspace N. It is not necessary that the distribution of $E_1$, $E_2$, etc., is a uniform distribution. Only the realizations of sequences $E_1$, $E_2$, … must be such that the distribution of the occurrences of a state in N and C are the same under different circumstances. Of course, if sequence $E_1$, $E_2$, always has the same distribution function, the distribution of the occurrence times of a state in N would be the same. Then, the inter-arrival times of states in N would be exponentially distributed since that is a rare event.

If the software states form a random walk in the space N u C, then a state in N is achieved as a random peak, which usually occurs with exponentially distributed inter-arrival times.

In order to come to a result based on the complicated structure of the software, we assume the following:

- The software failures are decoupled so that we can assume that they can be considered as being independent of each other.
- The fraction of failures in the software is small so that they should occur only rarely.

When the software is then subject to exterior influence, it could be possible to model the occurrence of each of the N software errors with one counting process, say $X_i(t)$. This process counts the failures of the software and is a renewal process, e.g. see [10] for the mathematical explanation of a renewal process. In fact, it means that the inter-arrival times of a software failure would have an arbitrary distribution and they would be independent. This choice can be motivated by the two assumptions above.

The overall counting process of all software errors (they show up as failures) is then

$X(t) = \Sigma X_i(t)$,

where the total runs over i. Now, we can apply Grigelionis' theorem [12].

If $E(X_i) = t/a_i$

$\Sigma 1/a_i \rightarrow \lambda$ for any fixed t

$\sup(X_i(B) >= 1) \rightarrow 0$ as $n \rightarrow \infty$ for every bounded set B of the time interval [0, ∞),

then X(t) converges to a homogeneous Poisson process with intensity $\lambda$.

This motivates many of the old software models using Poisson processes, see [9], Section 10. They are homogeneous Poisson processes if there are no changes and then all inter-arrival times of the states in N are exponentially distributed. If the software changes, e.g. improves, the occurrence rate changes and a non-homogeneous Poisson process is used.

What does Grigelionis' theorem mean practically?

- The software needs to be subject to an exterior influence that can be considered as random so that the software errors are triggered randomly.
- The occurrence probability of a single software error must be sufficiently small and the occurrence of different software errors must be considered as independent.
- The process of triggering software errors must be stationary.
- If a software failure occurs, the system is restarted and operates further on.

Note: If the software is changed, the stationarity assumption does not hold any more. It is then possible to find other models for reliability growth based, for example, on a non-homogeneous Poisson process, e.g. see [9].

Several variants of such limit theorems exist [18], e.g. the classic Palm-Khintchine theorem. Besides these simple models, other more complicated models exist that take into account a certain structure of the software, e.g. see [15]. Here, the software structure is modeled as a semi-Markov process, with software modules taking the role of states and control being transferred between the software modules. It is assumed that there may be failures within the modules with failure processes being a Poisson process. Also, the transfer of control may be subject to failure. The resulting process is quite complex, but again the limiting process is a Poisson process. The major requirement on which the result is based is that the modules and the transfer of control are very reliable, meaning that failures are rare and the failure rates of both processes are much lower than the switching rate.

We will now treat systematic hardware failures in the same manner as software failures. A systematic hardware failure is always present in the hardware and waits until its activation by an exterior trigger. This is comparable with the activation of a software error. However, it needs to be taken into account that, after the occurrence of a hardware fault, the system might fail and cannot be restored very easily to a working state. If it is repaired, comparable arguments as for the software might hold.

For many proven-in-use arguments, exponential distribution is used. This is then motivated by the Poisson process describing the failure counts of the systems under study.

It is necessary to distinguish between two aspects here.

A Poisson process is a failure counting process of a repairable system. That means a system is "repaired", e.g. by restarting it after a software failure, and, since influence on the system is random and the detected software error is activated seldom, we can just neglect the effect of it being activated a second time. After the occurrence of a hardware failure, the system is minimally repaired, i.e. just the failed item is replaced.

The assumption of stationarity, however, does not allow aging to be used in the model.

The Poisson process has the property that the inter-arrival times are exponentially distributed, which makes the use of statistical models very easy.

Note: It is necessary to distinguish the Poisson counting process from an exponential distribution that is used for the lifetime statistics of a non-repairable system, e.g. see [13].

Now, a short note is provided on how to take into account possible internal sources of random influence as intentionally or unintentionally introduced random number generators. They can be treated in the set-up as described above in the same manner as the exterior random influences.

## Examples

The Therac-25 was a radiation therapy machine which was involved in at least six incidents between 1985 and 1987, in which patients were given massive overdoses of radiation. The operational experience as reported by [6] started in 1982 with a total of 11 units. The details have not been reported, but for the example we assume that each unit was in operation on average for less than half the period between 1983 and 1987, say in total 20 unit-years. Assuming 250 operating days per year and 20 patients per day, we might assume 100,000 individual cases of treatment. For the sake of simplicity of the argument, let us assume that all qualitative requirements for proven-in-use arguments are fulfilled (which was probably not true in this case).

Let us assume we would pursue a proven-in-use argument on June 2, 1985, the day before the first incident, and that half the reported operational experience would have been gained. This would amount to about 50,000 flawless operating hours, which is an order of magnitude far removed from the minimum number of operating hours demanded for a SIL 1 claim [1] (300,000 operating hours without dangerous failures would be required). For a comparison, let us regard the data collected until January 17, 1987. Here, we would have doubled the number of operating hours, but we had six failures and probably software modifications, so the results would be even worse.

Two particular software failures were reported as root causes of the incidents [6]:

1. One failure only occurred when a particular non-standard sequence of keystrokes was entered on the terminal … within eight seconds. This sequence of keystrokes was improbable… it took some practice before operators were able to work quickly enough to trigger this failure mode.
2. In a second failure scenario, the software set a flag variable by incrementing it, rather than by setting it to a fixed non-zero value. Occasionally, an arithmetic overflow occurred, causing the flag to return to zero and the software to bypass safety checks.

These scenarios are similar in that there was a fault in the software from the very start, which was only triggered under a very particular environmental condition that is more or less of a random nature. So, it is reasonable to assume that the overall system behavior appeared randomly to an outsider, as it was triggered by an operational condition (e.g. particular treatment or particular set of parameters) and an error condition in the software which was triggered only with a particular probability (e.g. probability that the flag returned to 0). So, the overall failure behavior could be described by a possibly time-dependent failure rate $\lambda(t)$.

Ariane 4 was an expendable launch system which had 113 successful launches between 1988 and 2003 (and three failures) [17]. Its successor's test flight, Ariane 5 flight 501, failed on June 4, 1996, because of a malfunction in the control software. A data conversion caused a processor trap (operand error). The software was originally written for Ariane 4 where efficiency considerations (the computer running the software had an 80% maximum workload requirement) led to four variables being protected with a handler while three others, including the horizontal bias variable, were left unprotected because it was thought that they were "physically limited or that there was a large margin of error". The software, written in Ada, was included in Ariane 5 through the reuse of an entire Ariane 4 subsystem. As a result of a qualitative proven-in-use argument, the extent of testing and validation effort was reduced. Ariane 5 reused the inertial reference platform, but Ariane 5's flight path differed considerably from that of Ariane 4. Specifically, Ariane 5's greater horizontal acceleration caused the computers on both the back-up and primary platforms to crash and emit diagnostic data misinterpreted by the autopilot as spurious position and velocity data.

Let us assume that the component worked error-free in all 116 missions (the three losses had other causes) and that a mission would last on average one hour. It is clear that this operational experience is far removed from any reasonable SIL claim but it is interesting to note that, in this example, the change in operational environment would have invalidated any argument based on operational experience. In addition, the component was integrated into a different system, which would also have to be argued as sufficiently similar.

## Summary and conclusions

We have seen in the previous section that different and independent approaches lead to the same result, a limiting Poisson distribution, under very general requirements. And we are sure that, if we put in more work, we could even expand the results.

The situation is very similar to the central limit theorems in statistics. If there is no dominant influence or considerable dependence among the random variables, their normalized sum then tends to a normal limit distribution under very general assumptions. If some of the assumptions do not hold or we combine them with other functions, other limiting distributions would then occur, e.g. extreme value distributions.

So, we are back to the well-known Poisson assumptions, but we can relax them due to the limit theorem quoted above. It is well-known that the Poisson process arises when the following four assumptions hold:

1. Proportionality: The probability of observing a single event over a small interval is approximately proportional to the size of that interval.
2. Singularity: The probability of two events occurring in the same narrow interval is negligible.
3. Homogeneity: The probability of an event within a certain interval does not change over different intervals.
4. Independence: The probability of an event in one interval is independent of the probability of an event in any other non-overlapping interval.

For a particular application, it is sufficient that these assumptions hold approximately and are plausible. The singularity requirement needs no discussion; it will be fulfilled in any system with reasonable operational experience. The proportionality and homogeneity assumptions may be weakened, still resulting in a Poisson process, as the limit theorems hold for even the superposition of general renewal processes. However, the components must be well tested and be reasonably error-free as the limit theorems demand rare failure events. The independence and the stationarity requirements are more restrictive. They mean in particular that the environment must not change significantly and that the components regarded should be very similar. In particular, the Ariane 4 example has illustrated the problem concerning changes in the environment. And Therac 25 has shown that, when the system itself is changed, the operational experience may be very limited and may not even be sufficient to claim SIL 1. In both examples, we have assumed a working complaint management system which would be another issue to be discussed.

Since many standards (e.g. see [1]) and approaches use the Poisson process and the assumptions stated in the section above, we can now specify the requirements that must be necessarily fulfilled to apply such a proven-in-use approach:

- The main requirement is the random influence of the environment on the component and that it must be representative. This means that a component which is used for a proven-in-use argument must be in a typical environment for this type of component and the environment must be such that the influences are typical for this type of use, including all the changes in it. So, all components must be operated at least in similar environments. And all components compared must be similar. In particular, this means that, if bugs are fixed, the component has usually to be regarded as a different component.
- All failures must be recorded and it must be possible to observe them.
- All lifetimes and times of use must be recorded. It must also be recorded when a unit is out of service for a certain time interval and also when it is not used any more.

All these three points are nothing else than the requirements for a good statistical sample as can be found in good statistical textbooks, e.g. [14], and they do not differ from earlier results such as [3] or [15].

It is very interesting that the conclusion was already known in 1975 [16]: "This amounts to giving *carte blanche* to the reliability engineer to assume a Poisson distribution in such situations". And also another conclusion is still valid: "… it is common practice to make this assumption willy-nilly – the results contained here simply confirm the engineer's intuition". We wholeheartedly agree with this conclusion and so should also standardization committees, making life for engineers as easy as possible with minimum requirements, but having the assessors ensure that these requirements are really fulfilled. That this may still not be an easy task is finally illustrated by examples.